\documentclass[twocolumn]{article}

\usepackage{PRIMEarxiv}

\usepackage[utf8]{inputenc} 
\usepackage[T1]{fontenc}    
\usepackage{hyperref}       
\usepackage{url}            
\usepackage{booktabs}       
\usepackage{amsfonts}       
\usepackage{bm}            
\usepackage{nicefrac}       
\usepackage{microtype}      
\usepackage{lipsum}
\usepackage{fancyhdr}       
\usepackage{graphicx}       

\usepackage{amsmath,amssymb,amsfonts}
\usepackage{algorithmic}
\usepackage{graphicx}
\usepackage{textcomp}
\usepackage{mathptmx} 
\usepackage{multirow}
\usepackage{enumerate}
\usepackage{color,soul}
\sethlcolor{yellow}

\usepackage{times}
\usepackage{subcaption}

\usepackage[numbers]{natbib}
\usepackage{booktabs}
\usepackage{lscape}
\usepackage{comment}
\usepackage{verbatim}
\usepackage{ltablex}

\usepackage{stfloats}
\usepackage{lscape}

\usepackage{wrapfig}

\usepackage{longtable}

\usepackage{varwidth}
\usepackage{xcolor, colortbl}

\usepackage{enumitem}

\usepackage{orcidlink}

\title{Privacy-Preserving Explainable AIoT Application via SHAP Entropy Regularization }
\author{
 Dilli Prasad Sharma {\orcidlink{0000-0001-5497-3850}}, Xiaowei Sun, Liang Xue  \\
  School of Information Technology, York University, Toronto, Ontario, Canada \\
  \texttt{\{dilli,  xiaoweis, lxue03\}@yorku.ca}  \\
  \and
  \textbf{Xiaodong Lin} \\
  School of Computer Science, University of Guelph, Ontario, Canada \\ 
  \texttt{xlin08@uoguelph.ca} \\
  \and
 \textbf{Pulei Xiong} \\
  National Research Council of Canada, Ottawa, Ontario, Canada \\
  \texttt{pulei.xiong@nrc-cnrc.gc.ca}   
}
\begin{document}
\twocolumn[ {%
\begin{@twocolumnfalse}
\maketitle
\begin{abstract}

The widespread integration of Artificial Intelligence of Things (AIoT) in smart home environments has amplified the demand for transparent and interpretable machine learning models. To foster user trust and comply with emerging regulatory frameworks, the Explainable AI (XAI) methods, particularly post-hoc techniques such as SHapley Additive exPlanations (SHAP), and Local Interpretable Model-Agnostic Explanations (LIME), are widely employed to elucidate model behavior. However, recent studies have shown that these explanation methods can inadvertently expose sensitive user attributes and behavioral patterns, thereby introducing new privacy risks. To address these concerns, we propose a novel privacy-preserving approach based on SHAP entropy regularization to mitigate privacy leakage in explainable AIoT applications. Our method incorporates an entropy-based regularization objective that penalizes low-entropy SHAP attribution distributions during training, promoting a more uniform spread of feature contributions. To evaluate the effectiveness of our approach, we developed a suite of SHAP-based privacy attacks that strategically leverage model explanation outputs to infer sensitive information. We validate our method through comparative evaluations using these attacks alongside utility metrics on benchmark smart home energy consumption datasets. Experimental results demonstrate that SHAP entropy regularization substantially reduces privacy leakage compared to baseline models, while maintaining high predictive accuracy and faithful explanation fidelity. This work contributes to the development of privacy-preserving explainable AI techniques for secure and trustworthy AIoT applications.
\end{abstract}
\end{@twocolumnfalse} 
}
]
------------------
\renewcommand{\thefootnote}{} 
\footnotetext[0]{%

\vspace{2mm}
This work has been submitted to the IEEE for possible publication. Copyright may be transferred without notice, after which this version may no longer be accessible.
}
\renewcommand{\thefootnote}{\arabic{footnote}} 

\section{Introduction} \label{sec:introduction}

The growing adoption of Artificial Intelligence of Things (AIoT) in smart home environments has enabled intelligent systems for energy management, activity monitoring,  and personalized automation~\cite{Siam2025Artificial, Das2023Explainable}. These AIoT-based systems continuously use elements of the AIoT ecosystem to collect data, train models, and anticipate user behavior to improve their efficiency and decision-making. For instance, smart energy management systems rely on predictive models that utilize fine-grained data from smart home appliances (e.g., fridge-freezers, kettles, dishwashers) to forecast consumption patterns and support decisions such as load balancing, scheduling, and real-time optimization~\cite{Al2018Smart, Kong2017Short}. As these systems become more embedded in daily life, people depend on them more and more. This increasing reliance underscores the importance of ensuring that these systems are safe, trustworthy, reliable, and accountable.

Explainable AI (XAI) describes AI models for ensuring transparency, interpretability, and trustworthiness in AI systems~\cite{Ullah2024Explainable}.  XAI methods such as SHapley Additive
exPlanations (SHAP)~\cite{Lundberg2017Unified} and Local Interpretable Model-Agnostic Explanations (LIME) ~\cite{Ribeiro2016WhyShould} have been widely adopted across various industries and standards regulatory frameworks, including EU AI Act \cite{EUAIAct2024}, ISO/IEC 23894 \cite{ISO2023}, OECD AI Principles \cite{OECD2019}, and NIST AI Risk Management Framework \cite{NIST2023AI} to make the AI system's decisions interpretable and transparent to end users and stakeholders. 

However, recent studies reveal that post-hoc XAI methods such as SHAP and LIME can inadvertently act as vectors for privacy leakage by revealing sensitive behavioral patterns, even no direct access to raw input data~\cite{Liu2024Please, Yan2023Explanation, Luo2022Feature, Quan2022Amplification, Shokri2021Privacy}. Privacy leakage risks are particularly acute in smart home AIoT applications such as energy consumption forecasting, where explainable AI techniques (e.g., SHAP or LIME) can unintentionally reveal sensitive personal information, including occupancy patterns, daily routines, sleep and cooking schedules, presence at home, or specific appliance usage. These fine-grained explanations can be misused for behavioral profiling, device identification, or reconstructing the original input data, raising serious concerns under data protection regulations such as GDPR~\cite{GDPR2016} and CCPA~\cite{CCPA2018}. This growing risk emphasizes the need for explainable AI methods that not only provide transparency but also safeguard user privacy and ensure compliance~\cite{Allana2025Towards}. In response, we introduce a privacy-preserving approach based on SHAP entropy regularization, which reduces privacy leakage risks by enforcing higher entropy in SHAP attribution distributions during training. Our method limits the exposure of behavioral signals by reducing dependence on a few high-contributing features, thereby directly minimizing feature-specific privacy risks. The {\bf key contributions} of this work are summarized as follows:
\begin{itemize}
    \item Proposed a novel privacy-preserving explainable AI approach using SHAP entropy regularization for AIoT smart home applications. The SHAP entropy regularization method penalizes concentrated, low-entropy feature attributions by promoting a more even distribution of feature importance, thereby making explanations harder to associate with individuals and reducing privacy risks.    
    \item Developed a SHAP entropy-regularized Long Short-Term Memory (LSTM) regression model to effectively implement our approach on sequential smart home energy data. This model effectively captures temporal dependencies in appliance-level consumption patterns while incorporating privacy-preserving explanation regularization.
    
    \item Designed a suite of SHAP-based privacy attacks, including SHAP entropy attack, membership similarity attack, divergence attack, rank correlation attack, and rank consistency attack. These diverse attacks provide a comprehensive framework for assessing privacy leakage across various aspects of explanation behavior, including attribution stability, distributional divergence, and feature rank consistency.
    
    \item Conducted extensive experiments on an appliance-level smart home energy consumption dataset to validate the effectiveness of our proposed SHAP entropy regularization. Comparative evaluation against standard LSTM (BaselineLSTM) and differential privacy-enabled LSTM (DP-LSTM) models using SHAP-based privacy attacks and utility metrics shows that our method consistently outperforms both in privacy preservation and predictive accuracy.
    

    
\end{itemize}

The rest of the paper is organized as follows: Section \ref{sec:related-work} reviews the related work. Section \ref{sec:threat-model} discusses the threat model, including adversary assumptions and attacks considered. Section \ref{sec:proposed-approach} describes our proposed approach and model structure. Section \ref{sec:experimental-evaluation} discusses the used dataset, experimental setup, evaluation, and results discussion. Lastly, Section \ref{sec:conclusion} concludes this work and suggests future research directions.

\section{Related Work} \label{sec:related-work}
The increasing integration of explainable AI/ML into IoT-based smart applications has brought privacy risks to the forefront, particularly in smart homes, healthcare, and industrial IoT systems~\cite{Dwivedi2023Explainable, Aouedi2024Survey}. Several privacy-preserving techniques, including differential privacy (DP), federated learning (FL), homomorphic encryption (HE), secure multiparty computation (SMPC), and anonymization approaches, have been researched to protect user privacy in these AI-powered applications. We provide a brief overview of these key privacy-preserving methods alongside recent advancements in energy consumption forecasting. 

Several studies have explored the integration of differential privacy (DP) into explainable AI frameworks to enhance user confidentiality while maintaining model interpretability and utility. \citet{Harder2020Interpretable} discussed interpretable and differentially private predictions by analyzing the trade-off between interpretability, privacy, and accuracy. They introduced methods for generating differentially private (DP) local and global explanations in classification tasks. Similarly, \citet{Suriyakumar2021Chasing} applied DP methods in clinical prediction tasks, offering practical insights into privacy-utility trade-offs in sensitive healthcare data. They also highlighted that DP mechanisms can disproportionately impact certain demographic groups, potentially introducing bias. \citet{Huang2023Accurate} propose Laplace and DP-recourse methods that add DP noise to create useful, private counterfactual explanations. While innovative, their effectiveness depends on balancing noise and fidelity, which may limit use in high-stakes settings needing precise counterfactuals.

Recent advancements in FL have focused on unifying privacy, interpretability, and adaptability. \citet{Namakshenas2024IP2FL} proposed an interpretation-based privacy-preserving FL framework which integrates additive homomorphic encryption and Shapley values to enhance privacy and explainability in industrial cyber-physical systems. Similarly, \citet{Alzamil2025Federated} presented a transformer-based FL framework for electricity forecasting that ensures interpretability through layer-wise attention maps and robustness via adaptive optimization. \citet{Bogdanova2023DC-SHAP} developed an explainable data collaboration framework that combines KernelSHAP with privacy-preserving distributed learning. Together, these approaches demonstrate the growing focus on unifying privacy, interpretability, and efficiency in decentralized AI systems.

SMPC and anonymization techniques have also been explored to enhance the privacy of XAI methods. \citet{Jetchev2023XorSHAP} introduced the first practical privacy-preserving algorithm for computing Shapley values of decision tree ensemble models under a semi-honest SMPC setting with full threshold security. \citet{Goethals2023Privacy} investigated explanation linkage attacks that exploit instance-based counterfactuals and proposed k-anonymous counterfactual explanations as a mitigation strategy. They also introduced a pureness metric to evaluate these explanations, showing that anonymizing the explanations can enhance both privacy and interpretability.

Advancements in energy consumption forecasting and model interpretability have significantly accelerated the integration of AI and ML in smart home applications. Briggs et al.~\cite{Briggs2020Privacy} addressed the critical need for privacy-preserving, user-centric explainable AI in smart home energy forecasting, demonstrating that FL facilitates accurate demand prediction while safeguarding raw data privacy. \citet{Shajalal2022Towards} proposed an explainable forecasting framework that integrates SHAP and Deep-LIFT~\cite{Li2021Deep} with an LSTM model, enhancing user trust through transparent and interpretable predictions. \citet{Bhandary2024Enhancing} performed a comparative evaluation of various ML and deep learning models for household energy forecasting, employing LIME and SHAP to provide robust and reliable explanations. Furthermore, \citet{Munir2024Energy} proposed an energy consumption prediction model using a light gradient-boosting machine (LightGBM) combined with explainable AI techniques, which was validated on household datasets and benchmarked against existing methods. \citet{Zhao2015Consumer} presented an analytical methodology to extract and classify key demand patterns from smart meter data across four daily periods, including overnight, breakfast, daytime, and evening, revealing peak demand behaviors closely associated with time-of-day and seasonal variations. However, most of these forecasting and explanation methods focus on accuracy and interpretability, without considering the privacy risks in their explanations~\cite{Shokri2021Privacy}.

Despite recent advances in explainable AI (XAI) and privacy-preserving techniques, current state-of-the-art approaches still exhibit several limitations. First, most existing privacy-preserving methods in XAI—such as DP, FL, and SMPC focus on protecting raw data or model parameters but do not address the privacy risks associated with the explanation outputs.  Second, none of the existing studies investigates explanation-based regularization approaches. Third, there is a lack of research on membership inference attacks targeting explanation mechanisms (e.g., SHAP values), specifically in AIoT applications. To address these gaps, we propose a novel privacy-preserving method that introduces SHAP entropy regularization into the model training process. This approach aims to reduce the privacy leakage risk associated with explanation outputs while preserving predictive accuracy. To the best of our knowledge, this is the first work to employ explanation-based regularization to mitigate membership inference attacks targeting SHAP values in AIoT applications.

\section{Threat Model} \label{sec:threat-model}
In this section, we present the adversary model and SHAP-based privacy attacks. The adversary model outlines the adversary’s capabilities, assumptions, objectives, and the attack types considered. We also introduce a suite of SHAP-based privacy attack methods with explanation outputs.

\subsection{Adversary Model} \label{subsec:adverasary-model}
In this work, we adopt a gray-box (semi-white-box) threat model that realistically captures the privacy risks associated with explainable smart home energy forecasting systems, where explanation outputs (SHAP values) may be exposed to users, vendors, or third-party applications for transparency, interpretability, trust, and regulatory compliance. These explanation outputs can be accessed via dashboards, direct/indirect sharing, or APIs. We assume an adversary has moderate domain knowledge (e.g., typical appliance usage patterns, household energy trends) and partial information about the input feature space or its distribution. In this gray-box setting, the adversary observes the SHAP values corresponding to some input instances but does not have access to model parameters or training data.

An adversary aims to uncover sensitive information by analyzing patterns in model explanations and potentially launching privacy attacks~\cite{Liu2024Please}. To systematically evaluate the privacy risks posed by SHAP-based explanations, we consider three primary inference attacks to assess privacy risks in our setting:
\begin{itemize}
    \item Membership inference attack~\cite{Shokri2017Membership}: In this attack, an adversary attempts to determine whether a specific household’s consumption data was part of the model’s training set, risking individual privacy exposure.
    \item Property inference attack~\cite{Ganju2018Property}: This attack targets the extraction of global statistical characteristics of the training data, such as dominant appliance usage patterns or occupancy trends, revealing sensitive aggregate information.
    \item Explanation memorization attack~\cite{Li2024Privacy}: It exploits unique or memorized explanation patterns to identify specific training samples included during model training.
\end{itemize} 

Based on these inference attack scenarios, we further define five complementary SHAP-based attacks to enhance contextual relevance and provide a more comprehensive privacy assessment. These are discussed in the following section.


\subsection{Privacy Attacks in SHAP Explanations} \label{subsec:privacy-attacks-in-shap}
In this section, we present five inference-related privacy attacks on SHAP explanations based on the adversary model discussed earlier. These attacks use statistical and information-theoretic patterns to analyze and reveal private sensitive information via SHAP explanations. The privacy inference attacks proposed in this work are designed to exploit the information encoded in the SHAP attribution vector. 

Let $f: \mathbb{R}^d \rightarrow \mathbb{R}$ be a energy forecasting model trained on appliance-level consumption data $\mathbf{x}_i \in \mathbb{R}^d$, with prediction target $y_i$, and let  $\boldsymbol{\phi}(x) = [\phi_1(x), \phi_2(x), \ldots, \phi_d(x)]$ denote the SHAP attribution vector for an input $x$, where $d$ is the total number of features. Using this model and notation, we define our proposed SHAP-based privacy attacks as follows: 

\subsubsection{SHAP Entropy Attack} \label{subsec:shap-entropy}
In this attack, an adversary seeks to identify memorized or outlier samples by analyzing the entropy of the SHAP explanations. The key assumption is that memorized training samples tend to produce highly concentrated SHAP attribution vectors, where only a few input features (e.g., specific appliances or time-of-day) dominate the prediction, resulting in lower entropy. This low-entropy distribution can be used as a signal for potential membership inference. To quantify this, first the SHAP values for each input sample are normalized to form a probability distribution over features, and then the SHAP entropy of input \( x \) is computed using Shannon’s Entropy ~\cite{Shannon1948Mathematical}, which are given by:
\begin{equation}
    \tilde{\phi}_i(x) = \frac{|\phi_i(x)|}{\sum_{j=1}^d |\phi_j(x)|}, \quad \text{for } i = 1, 2, \ldots, d
    \label{eq:normalized-shap}
\end{equation}
\begin{equation}
    \mathcal{H_{\text{SHAP}}}(x) = - \sum_{i=1}^d \tilde{\phi}_i(x) \log \tilde{\phi}_i(x)
    \label{eq:entropy}
\end{equation}
Lower SHAP entropy suggests the model relies on a few dominant features, often signaling memorization of training data. In our energy forecasting system, if SHAP explanations consistently focus on appliances like the kettle or oven extractor fan during winter evenings, low entropy may suggest that such a household profile was included in training data, potentially revealing private behavioral routines or appliance usages.

\subsubsection{SHAP Similarity Attack} \label{sec:shap-similarity}
In this attack, an adversary computes the pairwise similarity between SHAP explanation vectors using metrics such as cosine similarity or Euclidean distance to infer membership status. The cosine similarity between two SHAP vectors is defined as:
\begin{equation}
     sim(\phi(x), \phi(x')) = \frac{\phi(x) \cdot \phi(x')}{\|\phi(x)\| \, \|\phi(x')\|}
     \label{eq:cosine-similarity}
\end{equation}
where \( \phi(x) \) and \( \phi(x') \) denote the SHAP value vectors for input samples \( \phi(x) \in \mathcal{D}_{\text{test}} \) and  \(\phi(x') \in \mathcal{D}_{\text{train}} \), respectively.  A high similarity score indicates that \( x \) is likely to originate from the training set and enabling a membership inference attack. 

\subsubsection{SHAP Divergence Attack} \label{sec:divergence}

This attack aims to determine whether a target data sample was included in the training dataset by measuring the statistical similarity between SHAP value distributions. In this, a symmetric and bounded divergence is computed using Jensen-Shannon Divergence (JSD) that quantifies distributional similarity~\cite{Lin2002Divergence}. Let $P$ and $Q$ denote normalized SHAP distributions for the target and a reference instance, respectively. JSD is defined as:
\begin{equation}
    \text{JSD}(P \| Q) = \frac{1}{2} \text{KL}(P \| M) + \frac{1}{2} \text{KL}(Q \| M)
     \label{eq:jsd}
\end{equation}
where $M = \frac{1}{2}(P + Q)$, and $\text{KL}(\cdot \| \cdot)$ is the Kullback--Leibler divergence~\cite{Kullback1951Information}. A low JSD suggests that the target SHAP distribution aligns closely with training data, indicating membership inference. 

\subsubsection{ SHAP Rank Correlation Attack} \label{sec:shap-correlation}
In this attack, the adversaries analyze the correlation between the SHAP-based feature rankings of a target instance and known training reference samples. They leverage Spearman's rank correlation~\cite{Spearman1961Proof} to measure similarity in SHAP value ranking between a target instance \(x\) and training instance \(x'\). This rank correlation is defined as follows:
\begin{equation}
    \rho (x,x') = 1 - \frac{6 \sum_{i=1}^d (r_i - s_i)^2}{d(d^2 - 1)}
    \label{eq:rank-correlation}
\end{equation}
where $r_i$ and $s_i$ are the ranks of feature $i$ in the target and training reference vectors, respectively. A high correlation between SHAP value rankings indicates a preserved order of feature importance, suggesting potential vulnerability to both membership and property inference attacks. For example, if energy consumption at specific time intervals (e.g., 6–9 PM) exhibits a strong correlation with SHAP importance scores (e.g.,$\rho >0.95$), an adversary can infer with high confidence that the target sample was part of the training dataset. Similarly, these temporal SHAP patterns may be associated with specific user behavioral routines (e.g., cooking or heating), revealing sensitive lifestyle attributes and enabling property inference attacks.

\subsubsection{SHAP Rank Consistency Attack} \label{sec:kendal-tau}
Unlike the SHAP rank correlation attack, which assesses the strength of monotonic relationships of overall rank differences, the SHAP Rank Consistency Attack measures ordinal consistency between feature rankings across samples using Kendall’s Tau~\cite{Kendall1938New}. Kendall’s \( \tau \) computes the number of concordant and discordant pairs in SHAP rankings, and it is defined as folows:
\begin{equation}
    \tau(x, x') = \frac{C - D}{\frac{1}{2}d(d-1)}
    \label{eq:kendal-tau}
\end{equation}
where \( C \) and \( D \) are the counts of concordant and discordant pairs, respectively. A high \( \tau \) indicates a strong match in local feature importance orderings, enabling inference attacks.


\section{Proposed Privacy-Preserving Approach} \label{sec:proposed-approach}

In this section, we present our proposed approach that describes the SHAP entropy regularization method and the SHAP entropy regularized LSTM forecasting model.

\subsection{SHAP Entropy Regularization} \label{subsec:shap-entropy-regularization}
In this section, we present our proposed SHAP Entropy Regularization approach, designed to enhance privacy in explainable smart home energy forecasting models. The core idea is to encourage the model to generate SHAP explanations with higher entropy, resulting in more evenly distributed feature attributions. This helps prevent the model from over-relying on a small subset of features, which can lead to overfitting or memorized patterns that cause risk of exposing sensitive information and compromising user privacy. Incorporating regularization into the training objective is effective in reducing overfitting and improving generalization performance, particularly in deep learning models such as LSTM~\cite{Ghiasi2023Improving, Krogh1991Simple}. The training objective of the forecasting model incorporating SHAP entropy regularization is defined as follows:
\begin{equation}
\mathcal{L}_{\text{total}} = \mathcal{L}_{\text{mse}} + \lambda \cdot \mathcal{R}(\mathcal{H_{SHAP}}(x))
\label{eq:shap_reg_generic}
\end{equation}
where $\mathcal{L}_{\text{mse}}$ is the loss of the forecasting accuracy (it can me mean square error), $\lambda > 0$ is a hyperparameter controlling the regularization strength, $\mathcal{H_{SHAP}}(x)$ is a SHAP entropy computed using Eq.~\ref{eq:entropy}, and $\mathcal{R}(\cdot)$ is a penalty function designed to encourage SHAP entropy towards a desired target. A common instantiation for $\mathcal{R}(\cdot)$ is a quadratic penalty enforcing the entropy to be close to a threshold $\alpha$, and it is defined as:
\begin{equation}
    \mathcal{R}(\mathcal{H_{SHAP}}(x)) = \left(\alpha - \mathcal{H_{SHAP}}(x)\right)^2
    \label{eq:penalty}
\end{equation}
Using Eqs.~\ref{eq:shap_reg_generic} and ~\ref{eq:penalty}, we formulate the overall training objective of the forecasting model with SHAP entropy regularization as follows:
\begin{equation}
\mathcal{L}_{\text{total}} = \mathcal{L}_{\text{mse}} + \lambda \cdot \left(\alpha - \mathcal{H_{SHAP}}(x)\right)^2
\label{eq:shap_regu}
\end{equation}
This proposed SHAP-based regularization formulation in Eq.~\ref{eq:shap_regu} can be directly incorporated into the training pipeline of an explainable forecasting model. By penalizing low-entropy SHAP explanations, this approach encourages the model to generate more balanced and privacy-preserving feature attributions.

\subsection{SHAP Entropy Regularized LSTM Model} \label{subsec:model}

We design an LSTM-based regression model with SHAP entropy regularization and train it adaptively for time-series forecasting~\cite{Hochreiter1997Long, Ghiasi2023Improving}. An LSTM model is a good choice because our data is time-series and has temporal dependencies. Let the input to the model be a sequence \( \bm{X} \in \mathbb{R}^{B \times T \times D} \), where \( B \) is the batch size, \( T \) is the number of time steps, \( D \) is the input feature dimension. The model employs a multi-layer LSTM network with hidden size \( H \in \mathbb{N} \), number of layers \( L \in \mathbb{N} \), dropout probability \( p \in [0,1] \) between layers. At each time step \( t = 1, \dots, T \), and for each layer \( \ell = 1, \dots, L \), the LSTM updates the hidden and cell states as:
\begin{equation}
  \bm{h}_t^{(\ell)}, \bm{c}_t^{(\ell)} = \text{LSTM}^{(\ell)}\left( \bm{x}_t^{(\ell)}, \bm{h}_{t-1}^{(\ell)}, \bm{c}_{t-1}^{(\ell)} \right),
\end{equation}
where, input to the first layer is \( \bm{x}_t^{(1)} = \bm{x}_t \in \mathbb{R}^D \), and for higher layers is \( \bm{x}_t^{(\ell)} = \text{Dropout}\left( \bm{h}_t^{(\ell-1)} \right) \). The LSTM output across all time steps of the top layer is given as:
\begin{equation}
  \bm{H} = \left[ \bm{h}_1^{(L)}, \bm{h}_2^{(L)}, \dots, \bm{h}_T^{(L)} \right] \in \mathbb{R}^{T \times H}.
\end{equation}
The final prediction is based on the hidden state at the last time step, and given by:
\begin{equation}
  \hat{y} = \bm{w}^\top \bm{h}_T^{(L)} + b, \quad \text{where } \bm{w} \in \mathbb{R}^{H},\ b \in \mathbb{R}.
\end{equation}

We train the LSTM regression model using a loss function that combines prediction error and a SHAP entropy-based regularization as defined in Eq.~\ref{eq:shap_regu}. For each training batch \( B \), the mean squared error (MSE) and the SHAP entropy are computed using Eq.~\ref{eq:loss-per-batch} and Eq.~\ref{eq:entropy-per-batch}, respectively:
\begin{align}
    \mathcal{L}_{\text{mse}} &= \frac{1}{|B|} \sum_{i=1}^{|B|} \| \hat{y}_i - y_i \|^2 
    \label{eq:loss-per-batch} \\
    \mathcal{H}_{\text{SHAP}}(B) &= \frac{1}{|B|} \sum_{i=1}^{|B|} \mathcal{H}_{\text{SHAP}}(x_i)
    \label{eq:entropy-per-batch}
\end{align}
where, \( x_i \) and \( y_i \) denote the input–output pairs in batch \( B \), and \( \hat{y}_i = f_\theta(x_i) \) represents the model’s prediction for input \( x_i \), parameterized by \( \theta \). The total training loss for each batch is obtained by substituting the mean squared error \( \mathcal{L}_{\text{mse}} \), defined in Eq.~\ref{eq:loss-per-batch}, and the average SHAP entropy \( \mathcal{H}_{\text{SHAP}}(B) \), defined in Eq.~\ref{eq:entropy-per-batch}, into the regularized loss formulation in Eq.~\ref{eq:shap_regu}. The resulting batch-wise loss function is obtained as:
\begin{equation}
    \mathcal{L}_{\text{total}}(\theta) = \mathcal{L}_{\text{mse}} + \lambda \cdot \left( \alpha - \mathcal{H}_{\text{SHAP}}(B) \right)^2
    \label{eq:total_loss_batch}
\end{equation}

where \( \alpha \) denotes a target entropy threshold encouraging uniformly distributed feature attributions, and \( \lambda \geq 0 \) is a regularization weight. This training procedure is guided by a dual-objective optimization that not only minimizes predictive error through the MSE loss but also dynamically regularizes SHAP-based explanation entropy. Further, the training process can incorporate adaptive SHAP entropy regularization, where the regularization weight \(\lambda\) is dynamically adjusted based on model performance on a held-out validation dataset.

\begin{figure*} 
\centering
 {\includegraphics[width=0.9\textwidth,keepaspectratio]{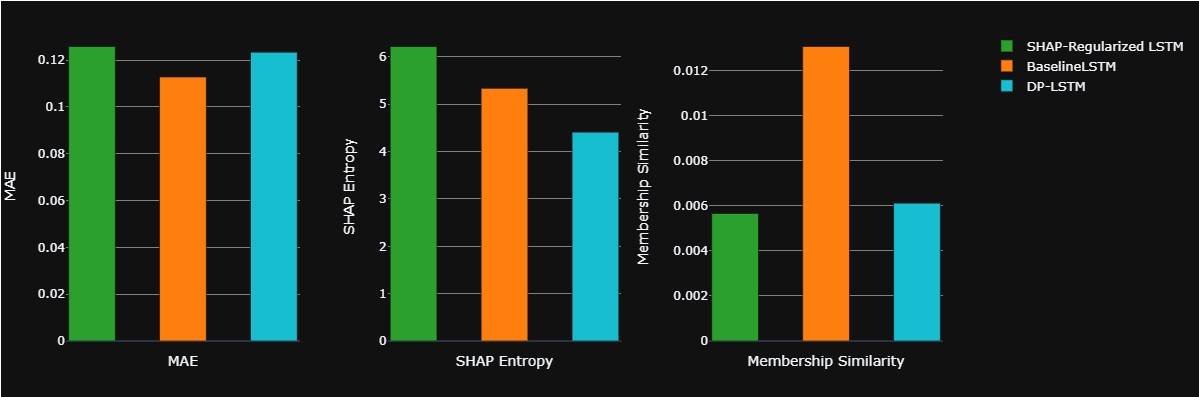}}
\caption {\small Comparing our SHAP-Regularized LSTM's Performance with Baseline LSTM and DP-LSTM Models.} 
\label{fig:comparing-models-performance}
\end{figure*}

\section{Experimental Evaluation \& Analysis} \label{sec:experimental-evaluation}
In this section, we describe the dataset used, detail the experimental setup, and discuss the evaluation results.
\subsection{Dataset} \label{subsec:dataset}
In this study, we use the REFIT Electrical Load Measurements dataset~\cite{Murray2015Data} to train and evaluate our proposed approach. The dataset contains power consumption records collected from 20 UK households between 2013 and 2015. Each house is equipped with 10 power sensors, including a current clamp that captures the aggregate household power usage and nine individual appliance monitors that record the active power consumption of selected appliances (e.g., televisions, computers, kettles, and washing machines) at the interval of approximately 6–8 second intervals. Data includes precise timestamps along with power readings for both individual appliances and the overall household consumption. For our experiments, we used data from three different houses and preprocessed them to compute hourly power consumption values. Table~\ref{tab:dataset-summary} presents a summary of the used data, including the time durations for training and testing splits, along with their corresponding sample sizes.
\begin{table}[ht]
\centering

\caption{Summary of data used in the experiments}
\resizebox{0.48\textwidth}{!}{%
\begin{tabular}{|l|l|l|l|}
\hline
\textbf{House} & \textbf{Time duration} & \textbf{Data split} & \textbf{Data size (samples)} \\
\hline
\multirow{2}{*}{House-1} 
& 10/9/2013 13:00 -- 3/3/2015 20:00 & Train & 12,248 \\
& 3/3/2015 21:00 -- 7/10/2015 11:00 & Test  & 3,063 \\
\hline
\multirow{2}{*}{House-2} 
& 9/17/2013 22:00 -- 1/24/2015 01:00 & Train & 11,836 \\
& 1/24/2015 02:00 -- 5/28/2015 08:00 & Test  & 2,959 \\
\hline
\multirow{2}{*}{House-3} 
& 9/25/2013 19:00 -- 1/29/2015 16:00 & Train & 11,782 \\
& 1/29/2015 17:00 -- 6/2/2015 10:00 & Test  & 2,946 \\
\hline
\end{tabular}
}
\label{tab:dataset-summary}
\end{table}

\subsection{Experimental Setup  } \label{subsec:experimental-setup}
We set up an experimental environment based on the threat model described in Section~\ref{sec:threat-model}, and conducted several experiments to validate the effectiveness of our proposed approach. We implemented an LSTM-based regression model to forecast the total power consumption of each house based on appliance data, as detailed in Section~\ref{subsec:model}. The model was trained separately for each of the three houses using the dataset described in Section~\ref{subsec:dataset}. For the comparative performance analysis, we trained the following three models and evaluated their effectiveness against both utility and the SHAP-based inference privacy attacks:
\begin{itemize}
   \item {\bf SHAP-Regularized LSTM}: It is an LSTM  regression model trained using our proposed SHAP entropy regularization method. Its effectiveness against SHAP-based inference attacks was evaluated and compared with baseline LSTM and  DP-LSTM models.
   \item {\bf Baseline LSTM}: It is a standard LSTM regression model trained without any defense mechanisms. This model provides a baseline to evaluate its susceptibility to SHAP-based inference attacks and its performance comparison with our proposed models.
   \item {\bf DP-LSTM}: It is an LSTM model trained with a Differential Privacy (DP)~\cite{Dwork2006Calibrating} mechanism using the Opacus library~\cite{Yousefpour2021Opacus}. This model serves as a benchmark privacy-preserving technique, enabling a comparative evaluation of our proposed approach with a well-established defense mechanism.
   \end{itemize}
   All experiments were conducted using the PyTorch framework and executed on a Tesla V100 GPU to ensure consistent and scalable model training and evaluation.

\subsection{Evaluation \& Analysis} \label{subsec:evaluation}
We compare the experimental results of our SHAP-regularized model against baseline LSTM and DP-LSTM models using the utility metric Mean Absolute Error (MAE) and SHAP-based privacy attacks, including SHAP entropy, similarity, divergence, rank correlation, and consistency to evaluate utility and privacy risks. 

\subsubsection{Comparative Analysis via SHAP Privacy Attacks}
Fig.~\ref{fig:comparing-models-performance} presents the effectiveness of our approach with baseline LSTM and DP-LSTM models on House-2 data using  MAE, SHAP entropy, and similarly attacks. Results show that our model outperforms both baselines, with only a slight overhead in utility performance. We evaluated our model on data from three houses against both baselines. Table~\ref{tab:models-comparision} presents the comparative results across all houses, attacks, MAE, and models. Higher SHAP entropy and divergence, along with lower similarity, rank correlation, and rank consistency, collectively reflect stronger privacy preservation against SHAP-based inference attacks. The results show that our SHAP-regularized model consistently outperforms both the baseline LSTM and DP-LSTM across all privacy metrics, with the exception of House 1’s similarity score, where the DP-LSTM performs slightly better. This improvement in privacy comes at the cost of only a minor utility overhead.

\begin{table*} [ht]
\centering
\caption{\small {Comparative performance of the SHAP-regularized model, baseline LSTM, and DP-LSTM across three houses}}
\resizebox{0.95\textwidth}{!}{%
\begin{tabular}{|l|l|l|l|l|l|l|l|}
\hline
\textbf{House} & \textbf{Model} & \textbf{SHAP Entropy} & \textbf{SHAP Similarity} & \textbf{JS Divergence} & \textbf{Rank Correlation} & \textbf{Rank Consistency} & \textbf{MAE} \\
\hline
\multirow{3}{*}{House-1} & SHAP-Regularized LSTM & 6.0271 & 0.1746 & 0.6785 & 0.2413 & 0.1630 & 0.1376 \\

                         & Baseline-LSTM         & 5.9864 & 0.2344 & 0.6444 & 0.2889 & 0.2007 & 0.1236 \\

                         & DP-LSTM               & 5.2699 & 0.1149 & 0.5269 & 0.3356 & 0.2653 & 0.1513 \\
\hline
\multirow{3}{*}{House-2} & SHAP-Regularized LSTM & 6.2199 & 0.0057 & 0.3984 & 0.5836 & 0.4368 & 0.1258 \\

                         & Baseline-LSTM         & 5.3376 & 0.0131 & 0.3254 & 0.7237 & 0.5639 & 0.1128 \\

                         & DP-LSTM               & 4.4115 & 0.0061 & 0.2703 & 0.7444 & 0.6914 & 0.1234 \\
\hline
\multirow{3}{*}{House-3} & SHAP-Regularized LSTM & 6.5992 & 0.0565 & 0.4173 & 0.3990 & 0.2896 & 0.1527 \\

                         & Baseline-LSTM         & 6.5809 & 0.0573 & 0.4324 & 0.4791 & 0.3474 & 0.1796 \\

                         & DP-LSTM               & 5.5652 & 0.1895 & 0.3703 & 0.7163 & 0.6056 & 0.1361 \\
\hline
\end{tabular}
}
\label{tab:models-comparision}
\end{table*}
\subsubsection{Comparative Analysis via SHAP Attribution} 
\begin{figure*}
 {\includegraphics[width=\textwidth,keepaspectratio]{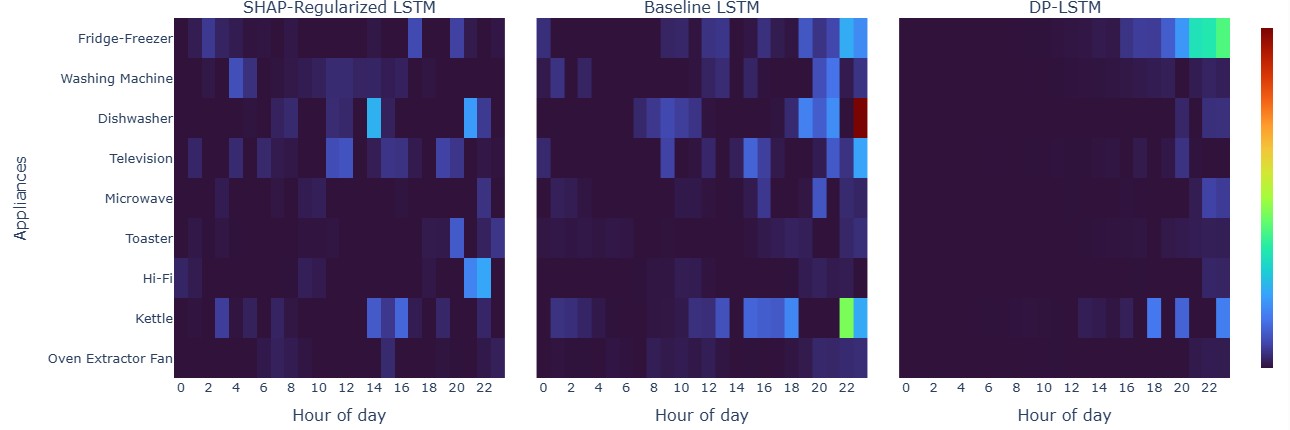}}
\caption {\small Comparing SHAP Attribution of  Appliances (Features) with three different models.}
\label{fig:shap-comparision}
\end{figure*}

Figure~\ref{fig:shap-comparision} presents the SHAP value heatmaps for the SHAP-Regularized LSTM, Baseline LSTM, and DP-LSTM models across different hours of the day and appliances. The SHAP-Regularized model produces diverse and less concentrated attributions, with patterns more evenly distributed over time and across appliances. This makes the explanations less predictable and indicates stronger privacy, as it becomes more difficult for an adversary to infer appliance usage. In contrast, the Baseline LSTM reveals strong and repeated attribution patterns for appliances such as Kettle, Fridge-Freezer, and Television during evening hours, which may expose sensitive user behavior. The DP-LSTM reduces some of these patterns, but its SHAP values are overly smooth and mostly near zero, except for a few late-hour spikes. This suggests the presence of strong noise from differential privacy, potentially degrading model utility. Overall, our SHAP-Regularized model offers improved privacy protection by obscuring consistent attribution patterns.

\subsubsection{Model Comparison via SHAP Entropy} 
\begin{figure*}
 {\includegraphics[width=\textwidth,keepaspectratio]{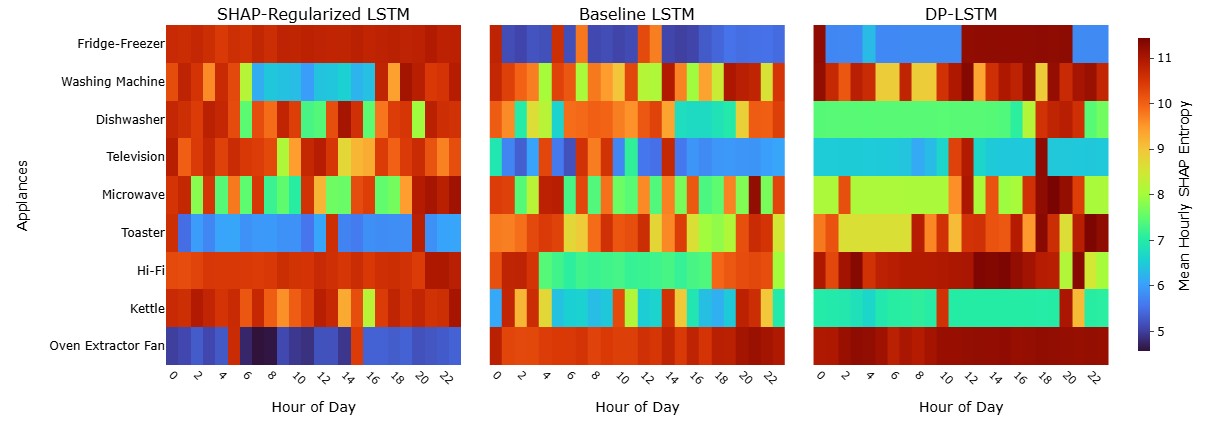}}
\caption {\small Comparing hourly SHAP entropy of appliances with three different models.}
\label{fig:entropy-comparision}
\end{figure*}

Fig.~\ref{fig:entropy-comparision} shows the hourly SHAP entropy distribution across various appliances (i.e., model features or predictors) for three different models: SHAP-Regularized LSTM, Baseline LSTM, and DP-LSTM. The SHAP-Regularized LSTM consistently exhibits higher entropy for the appliances, indicating more uniformly distributed feature attributions. This high entropy reflects reduced dependence on individual input features and thus enhances explanation-level privacy. However, some appliances, such as the oven extractor fan, toaster, and washing machine, still have lower entropy during specific daytime hours, likely due to their regular and predictable usage patterns. Despite these isolated vulnerabilities, the SHAP-Regularized LSTM shows strong potential in mitigating privacy risks by encouraging diverse explanations and suppressing concentrated attribution patterns.

However, the baseline LSTM demonstrates widespread low entropy for several appliances, including the fridge, television, microwave, and kettle, making it particularly vulnerable to inference attacks. This is because, without any regularization or privacy-enhancing mechanisms, the model tends to overfit to dominant features, producing highly concentrated SHAP values. The DP-LSTM, while designed to protect training data through noise injection, does not directly influence how explanations are formed. Consequently, it exhibits moderate entropy across many appliances, such as the dishwasher, television, and microwave, suggesting only partial mitigation of privacy risks with SHAP values.

\begin{figure*}[ht]
    \centering
    \begin{subfigure}[b]{0.48\textwidth}
        \centering
        \includegraphics[width=\textwidth,keepaspectratio]{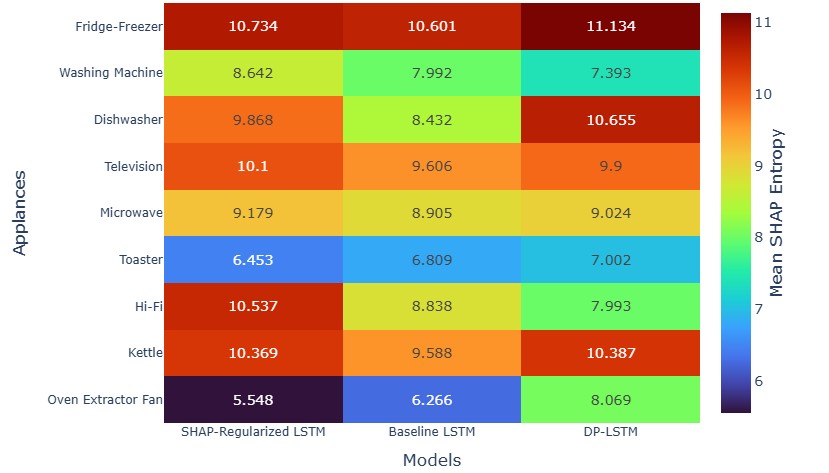}
        \caption{Aggregated SHAP Entropy}
    \end{subfigure}
    \hfill
    \begin{subfigure}[b]{0.48\textwidth}
        \centering
        \includegraphics[width=\textwidth,keepaspectratio]{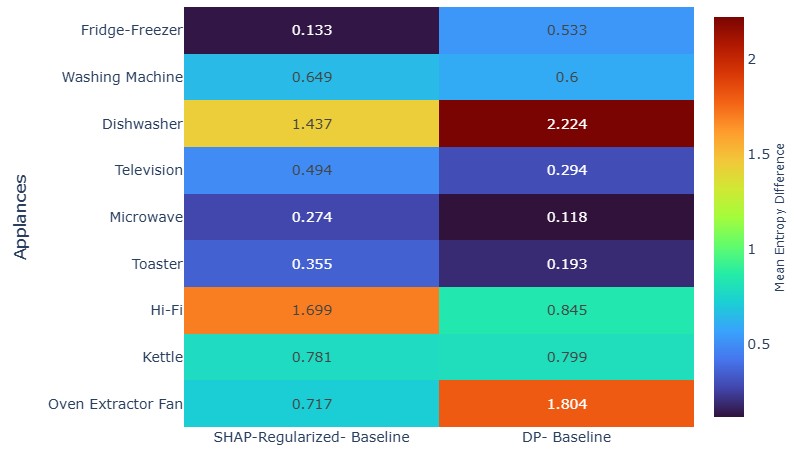}
        \caption{SHAP Entropy Difference}
    \end{subfigure}
    \caption{Comparing entropy of each appliance across models: (a) aggregate SHAP entropy, (b) their difference with baseline LSTM. }
    \label{fig:model-comparing-with-entropy}
\end{figure*}

We compare the entropy of each appliance’s SHAP attributions across models by analyzing the aggregated SHAP entropy and deviation from the baseline model. Fig.~\ref{fig:model-comparing-with-entropy} (a) and (b) present aggregate SHAP entropy comparison, and their differences relative to the baseline, respectively. This analysis helps identify the most vulnerable appliances under each model and provides an overall assessment of the privacy performance. The toaster consistently exhibits low entropy in all models, marking it as a persistently vulnerable feature. The oven extractor fan also remains relatively vulnerable, even under SHAP regularization. Most other appliances show higher entropy with the SHAP-Regularized LSTM, indicating stronger explanation privacy. While the DP-LSTM shows a modest improvement over the Baseline LSTM in terms of entropy, the difference is relatively small, indicating that differential privacy alone may not sufficiently enhance explanation-level protection.  These results indicate that SHAP Regularization provides the strongest defense among the three models by effectively increasing attribution diversity and reducing concentration on specific features.

\section{Conclusion} \label{sec:conclusion}
In this work, we propose a privacy-preserving explainable AI method for AIoT smart applications by integrating SHAP entropy regularization into an LSTM-based regression model. To rigorously assess explanation-level privacy risks, we develop a suite of SHAP-based privacy attacks, including SHAP entropy analysis, membership similarity, distributional divergence, rank correlation, and consistency measures. We validate our approach through extensive experiments on a benchmark smart home energy dataset, demonstrating that the SHAP entropy-regularized model significantly improves explanation privacy compared to both the standard baseline LSTM and DP-LSTM models. Overall, our method provides a practical and robust framework for preserving privacy in model explanations and feature attributions, thereby protecting sensitive user behavioral patterns in AIoT applications. Furthermore, it contributes to advancing trustworthiness in explainable AI methods through privacy-preserving methods, with broader relevance across both the IoT and AI research communities.

We plan to conduct the following {\bf future work} to further improve  this SHAP entropy regularization approach:
    \begin{itemize}
        \item Develop adaptive mechanisms to balance privacy and explanation fidelity~\cite{Nguyen2023XRAND} in real time, adjusting regularization strength based on context or data sensitivity.
        \item Extend our SHAP entropy regularization approach to multimodal AIoT data~\cite{Liu2024Privacy}, addressing privacy challenges in heterogeneous and cross-device environments.
        \item Develop a privacy risk assessment framework for explainable AI that integrates our SHAP-based privacy attacks with legal compliance~\cite{GDPR2016} and user understanding.
    \end{itemize}

\section*{Acknowledgment}
This project was supported by collaborative research funding from the National Research Council of Canada’s Artificial Intelligence for Logistics Program.

\bibliographystyle{IEEEtranSN}
\small{
\bibliography{references}
}
\end{document}